\documentclass{article}

\usepackage{authblk}
\usepackage{amsmath,amssymb}
\usepackage{bm}
\usepackage{graphicx}
\usepackage{array}
\usepackage{amssymb,dsfont,mathrsfs}
\usepackage[mathscr]{eucal}
\usepackage{subfigure}
\usepackage{verbatim}
\usepackage{wrapfig}
\usepackage{cite}
\usepackage{makeidx}
\usepackage{enumerate}
\usepackage{empheq}
\usepackage{float}
\usepackage{threeparttable}
\usepackage{caption}
\usepackage[left=25truemm,right=25truemm]{geometry}
\makeindex
\title{Summation of certain locally bilinear forms and its applications to the Fast Multipole Method}

\author{Yasuhiro Kajima\thanks{y-kajima@nifty.com, \ kajima@nzu.ac.jp}}
\affil{Nagoya Zokei University, Komaki, Aichi 485-8563, Japan}
\begin{document}
\maketitle

\

\

\begin{abstract}
{The Fast Multipole Method (FMM) reduces the computation of pairwise two-body interactions among $N$-particles to order $N$, whose computation cost should be of order $N^2$ by brute force. However, its implementation is somewhat complicated and requires a considerable amount of time to write the code. 
In this paper, I show a method that enables us to implement and write FMM algorithm code simply and briefly. FMM algorithm is composed of several steps. The main steps are Upward Pass and Downward Pass. Both the Upward Pass and Downward Pass include shift processes by which we move the centers of local expansions and multipole expansions. In this paper, I show a method that enables us to get rid of these processes.
As a result of this simplification, the coding of FMM becomes much easier, and we can save considerable computation time. I compared the accuracy and time required to calculate potential fields with that of the existing FMM code.}
\end{abstract}

\

\

\section{Introduction} 
The fast multipole algorithm developed by Greengard and Rokhlin \cite{101} enables us to calculate $N$-particles interaction of particles such as gravitational or 
electrostatic forces with $\mathcal{O}(N)$ operations with predictable error bounds and considered as one of the top 10 algorithms of the 20th century \cite{110}. 
Thus, a large amount of papers were published relating to this theory (such as \cite{120}-\cite{124}).
The interaction is divided into two parts; near interaction and the rest (not near) interaction. The near interaction part is the interaction of particles each two of which are within a given distance (or within some boxes). 
This is calculated directly with $\mathcal{O}(N)$ operations. 
The rest part is calculated by the FMM also with $\mathcal{O}(N)$ operations.
Thanks to the FMM algorithm, we can perform large-scale simulation (\cite{140}).
However, its implementation is not necessarily easy and 
the time needed to perform simulation is not necessarily short.
In \cite{140}, we spent about half the computation time for the computation of forces, 
even though we employed FMM.
Therefore, I propose a less time-consuming and easier to implement FMM in this paper.

\section{Theory}
\subsection{Greengard's Fast Multipole Method}
In the following, I describe a two-dimensional case, which is simpler than three-dimensional cases
but would be sufficient to understand more general cases intuitively. Unlike Greengard's FMM, our method for two-dimensional cases can be applied to three dimensional
cases straightforwardly. 
We use the following notation: 

\begin{enumerate}
\item "The (computation) box" is a (two-dimensional) square with sides of length five. We say the box is at level-zero. The size of the square can be chosen arbitrarily, but I set it five for convenience.
\item The box contains $N$-charged particles.
\item The box is composed of four same-size regular squares, and we call them level-one squares. 
Here, the "four" is because we are dealing with a two-dimensional case for simplicity. 
For a three-dimensional case,  it should be "nine".
\item Each level-one square is composed of four same squares, and we say the squares are at level-two. We repeat this procedure up to a given level $l$. If a square $m$ is at level $k$, we write 
$m$ as $m^k$ to indicate its level $k$ if necessary. $m^0$ is the computation box (Fig.1).

\begin{figure}[h]
\centering
\label{Fig.1}
\includegraphics[width=4cm]{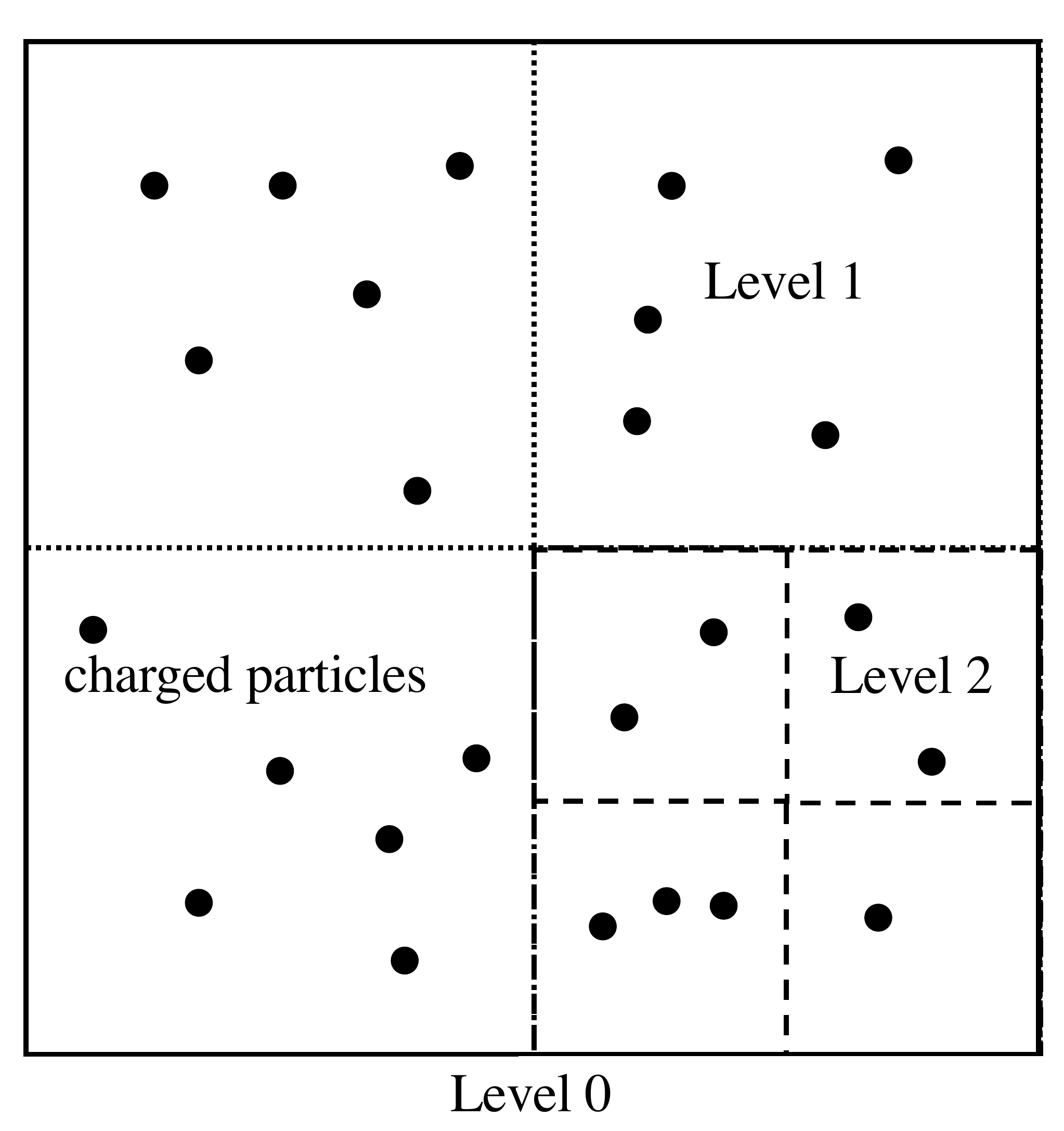}
\caption{Squares at level 0, 1, 2.}
\end{figure}

\item For $m_i^k$ with $k\ge2$, there is a list of squares around $m_i^k$ called 
"interaction list" (see Fig.2,\cite{101}). If $m_j$ is on the interaction list of $m_i$, we 
denote $m_j\in IL(m_i)$. We call same-sized squares well-separated if the distance between the squares
is $>$ length of the sides of them.
\end{enumerate} 

\begin{figure}[h]
\centering
\label{Fig.2}
\includegraphics[width=4cm]{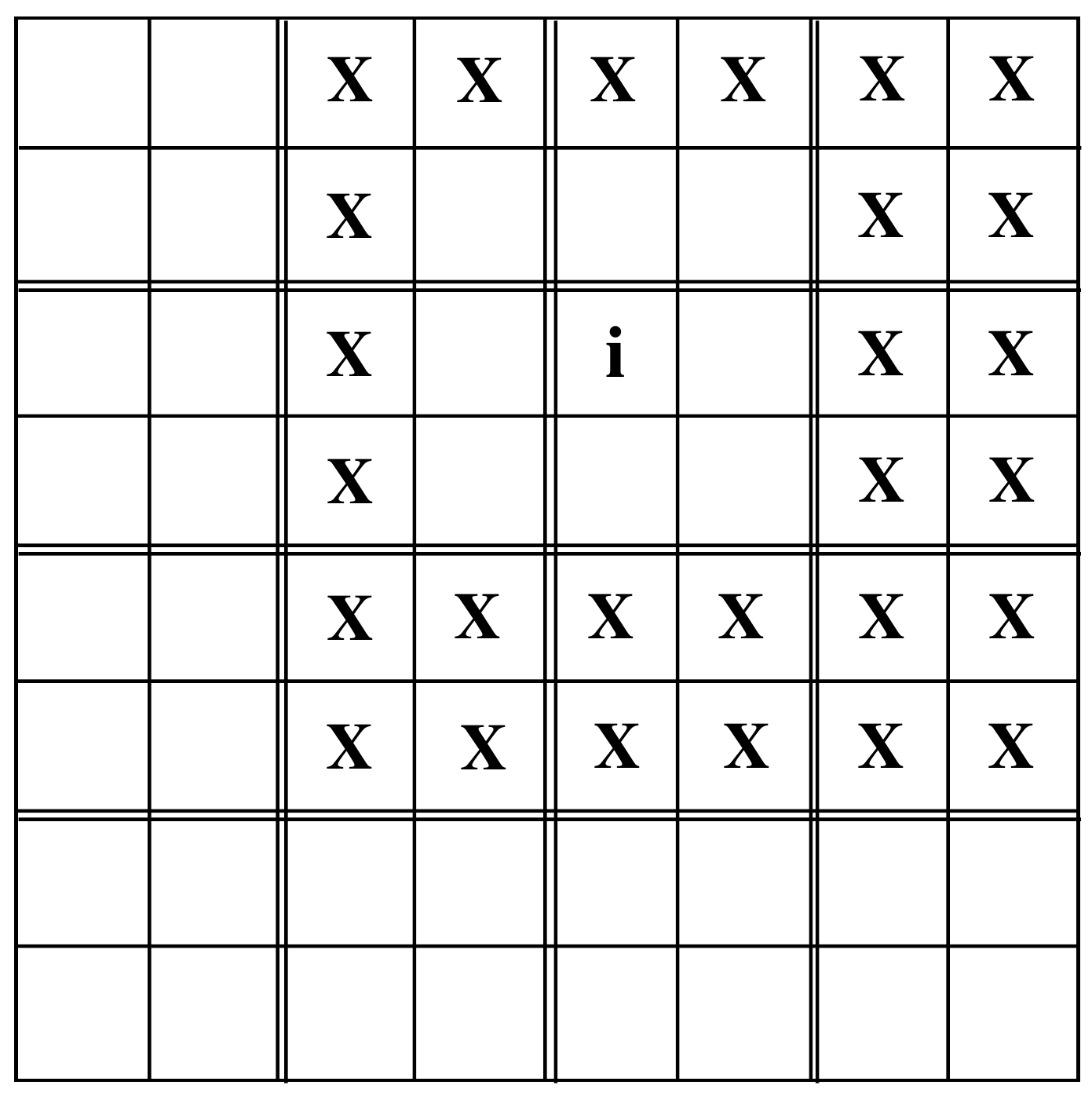}
\caption{Interaction list for square {\bf i}. We say that {\bf x}'s are in the interaction list of {\bf i}.}
\end{figure}

The method I describe in this paper (and Greengard's method as well) is not restricted to cases where computation boxes are regular squares, nor two-dimensional. We can apply our method to three-dimensional cases. However, for the sake of depicting figures, I restrict myself to two-dimensional cases. The method I describe here for two-dimensional cases can be straightforwardly applied to higher-dimensional cases.

Now, we recall some basic steps composing the Greengard's FMM briefly and then compare the steps of our
method with that. For details, please refer to \cite{101}. 
FMM is composed of the following steps: 
\\
\\
{\bf Upward Pass:} 
\begin{enumerate}[{U}-1]
\item Form multipole expansions $\Phi_{m^l_i}$ of the potential field due to charged particles in each finest level square $m_i^l$ about the square center using spherical harmonics (associated Legendre functions).
\item Form multipole expansions about the centers of all squares at all coarser levels. This step is done by shifting the multipole expansions $\Phi_{m^k_i}$ 
at level $k$ to the center of squares $m_j^{k-1}$ containing $m_i^k$. Then add the four resulting multipole expansions to obtain $\Phi_{m^{k-1}_j}$. Repeating this from $k=l$ to $k=1$, we obtain all $\Phi_{m^k_i}$'s.
The "shift" is a translation of associated Legendre functions.
\end{enumerate}
{\bf Downward Pass:}
\begin{enumerate}[{D}-1]
\item Form a local expansion $\Tilde{\Psi}_{m}$  
about the center of each square $m$ at each level $k<l$. 

This is done by first converting multipole expansions $\Phi_{m_i}$ 
for squares $m_i\in IL(m)$ to a local expansion about the center of $m$.
We denote this by $\Tilde{\Psi}_{m_,mi}$ 

Then, adding these multipole expansions $\Tilde{\Psi}_{m_,mi}$, we obtain $\Tilde{\Psi}_{m}$:

\[
\Tilde{\Psi}_{m}=\sum_{m_i \in IL(m)}\Tilde{\Psi}_{m_,mi}.
\]

\item Form a local expansion $\Psi_{m}$  and compute interactions at the finest level $l$. 

Let $\Psi_{m}=0$ for the square $m$ at level 0.
We shift the center of local expansion $\Psi_{m^k}$ of level $k$ to
the centers of $m_1$, $ m_2$, $m_3$, $ m_4$ in $m$, where they are at level $k+1$ 
(now, we have made four local expansions).
Then, we define $\Psi_{m_i^{k+1}}$ by adding these local expansions to
$\Tilde{\Psi}_{m_i^{k+1}}$. 
Repeating this procedure inductively, we obtain local expansions at the finest level $l$.

\end{enumerate}
{\bf Evaluate:}
\begin{enumerate}[{E}-1]
\item For each charge $c\in m_i^l$ evaluate the potential (or force) due to not near (not contained in $m_i^l$ nor adjacent eight squares $m_j^l$) particles.
\item Compute potential due to charges not contained in E-1 directly and add it to the results of E-1.
\end{enumerate}

\subsection{Our Method}
As mentioned above, the Greengard's Fast Multipole Method first evaluates the total action of charged particles in a square toward a charged particle in another same-sized square in its interaction list, and
sums up the total actions in a specific way so that the order of the calculation to be N. Greengard's FMM represented this total action using spherical harmonics. 
In our method, we represent total actions of charges in a square $m_i^k$ by charges fixed in position
in the square $m_i^k$.

We first describe our method in some general manner as
"summation of locally bilinear forms", and apply it to 
the computation of the potential of charges.
\subsubsection{Locally bilinear forms}
I describe here some results on "summation of locally bilinear forms".

We assume that there are $N$ points $c_i$ $(1\le i\le N)$ in the computation box $m^0$.
Each of these point $c_i$ is tagged to a different vector $v_i$ in
a $\mathtt{d}$-dimensional vector space $\bf{V}$.
We denote this relation by $c_i\rightarrow v_i$.
If $c_i\rightarrow v_i$ and $c_i\in m_k$, we write as $v_i\in m_k$ 
by abuse of language.
$v_i\in m^0$ always holds.
We will define the map explicitly for charges in two dimensional space later, where the image $v$ of $c$ is
determined by combining 
positions and electric charges of $c$.

We also assume that there are bilinear forms
$B_{m_i,m_j}(v_k,v_l)$ for $m_i\in IL(m_j)$.
They are bilinear with respect to vectors $v_k$, $v_l\in V$, however, 
the bilinear forms have specific meanings only for $v_k \in m_i$ and $v_l\in m_j$ and
vectors spanned by them (spanned by vectors in each square).
In this sense, we call them locally bilinear forms.

Then, we compute the summation of the locally bilinear forms for $v$ by
\[
F(v)=\sum_{v_k\in m_j, \ v\in m, \ 1\le k\le N}B_{m,m_j}(v,v_k).
\]
Note that $B_{m_i,m_j}(v_k,v_l)$ is not defined for $v_k$ near $v_l$ (i.e., in the inside
of adjacent or same square of $m_j$), we understand as $B_{m_i,m_j}(v_k,v_l)=0$ in this case.
We can compute interactions of well-separated
$N$ particles (i.e., interaction between points in well-separated squares) using $F(v)$ later.

The main result of this paper
is that we can compute the function $F(v)$ in a $\mathcal{O}(N)$ computation without "shift" procedure.
\\
\\
To show that the computation above is $\mathcal{O}(N)$, we proceed as follows: \\
\\
{\bf Upward Pass:} 
\begin{enumerate}[{$\rm\Bar{U}$}-1]
\item Form vectors $v_{m_i^l}$ for all $m_i^l$
 by $v_{m_i^l}=\sum_{v_k\in m_i^l}v_k$,
where $m_i^l$ are squares at level $l$ (the finest level).
\item Form vectors $v_{m_j}$ from finest squares to coarser squares
inductively by
\[
v_{m_j^k}=\sum_{{m_i^{k+1}}\subset m_j^k} v_{m_i^{k+1}}, \ \ (k+1\le l)
\]
\end{enumerate}
{\bf Downward Pass:}
\begin{enumerate}[{$\rm\Bar{D}$}-1]
\item We form locally linear forms $\Tilde{L}_m(v)$ for each square $m^k$ at each level $k<l$. 
For a square $m$, we add
$B_{m_i,m}(v_{m_i},v)$ for squares $m_i\in IL(m)$, $v\in m$ and denote the resulting 
bilinear form by $\Tilde{L}_m(v)$:
\[
\Tilde{L}_m(v)=\sum_{m_i \in IL(m)}B_{m_i,m}(v_{m_i},v).
\]
\item Compute interactions at the finest level $l$. 
Let $L_m(v)=0$ for the square $m$ at level 0.
We add the linear forms 
$L_{m^k}(v)$ of level $k$ to 
$\Tilde{L}_{m_i^{k+1}}(v)$'s ($i\in \{1,2,3,4\}$), where $m_1$,$ m_2$,$m_3$,$ m_4$ are the 
(level $k+1$) four squares in $m^k$.
We denote the resulting linear forms by $L_{m_i^{k+1}}(v)$.
Repeating this procedure inductively, we obtain the linear forms at the finest level $l$.
\end{enumerate}
For $v\in m^l$ (a finest level square), put $F(v)=L_{m^l}(v)$.
This is the same as previously defined $F(v)$.
Each of these additions (appeared in $\rm\Bar{D}$-1 and $\rm\Bar{D}$-2) is essentially 
addition of $\mathtt{d}$-dimensional dual vectors. 
The reason why this method is $\mathcal{O}(N)$ is almost the same as 
Greengard's FMM.
\\
\\

\subsubsection{Applications to FMM}
To apply the results obtained above, we proceed as follows.
For details, see the appendix.
\begin{enumerate}
\item First, we fix a one-to-one correspondence 
between a charged particle and $v\in \bf{V}$
\item Second, we construct bilinear forms $B_{m_i,m_j}(v_k,v_l)$ such that a interaction 
(such as Coulomb interaction) between
 $c_k\in m_i$ and $c_l\in m_j$ is equal to $B_{m_i,m_j}(v_k,v_l)$ 
 within a predictable error bound, where $c_k\rightarrow v_k$ and 
$c_l\rightarrow v_l$. 
In the following, an "error is predictable" means that the value $|$error$\times d^u |$ is bounded
where $d$ is the distance between the squares and we know the exponent $u$. $u$ becomes large if we increase 
the dimension $\mathtt{d}$. The error would be estimated more specifically, but we 
do not pursue this issue.
\item Lastly, we compute the interaction from nearby particles (which are not well-separated) directly.
\end{enumerate}

Note: The map $\rightarrow$ 
does not depend on the choice of squares containing a charge $c$, i.e.,
if $c\in m_1$ and $c\in m_2$ (here the level of $m_1$ and $m_2$ should be different),
the image of the map is the same. 
Because of this fact, we can get rid of "shift"'s in Upward Pass and Downward Pass.\\
\\

\section{Results}
I have prepared a Fortran program by modifying existing FMM program 
developed by Prof. Shuji Ogata (FMMP\cite{130}) to test execution time and accuracy. 
In the computation, I fixed $4\times4\times4=64$ positions in each cube 
precisely the same as Fig.2 in the appendix (The lengths of the sides are five.
We are now dealing with 
three-dimensional cases, so I use a cube instead of a box). The computations were performed 
in double precision, compiled with Intel Fortran compiler version 19.1, the CPU used is Core i7-9700.

Here, in this section, we compare the computation timings and accuracies of our program
with that of FMMP. In FMMP, the maximum order of multipoles is set to five except for a
test case shown in the last
row in Table 2 which is added to see how the maximum order of multipoles influences the results.
Our method can be applied to parallel computing as with Prof. Ogata's FMMP.
However, I have coded for one node this time.

In the following tests, I scattered charges randomly in a cubic box. The charges are $+1$ or $-1$.

\begin{table}[H]
\caption{\label{tab:table1}
Computation timings averaged over $10$ measurements for the case where\\
{\bf number of particles=80000, and maximum level $l$=3}}
\

\begin{threeparttable}[h]
\begin{tabular}{ccccccc}
\hline
program 
&\verb|t_up|\tnote{1}
&\verb|t_down|\tnote{2}
&\verb|t_pfs|\tnote{3}&total (sec)
&accuracy\tnote{4}\\
\hline
\hline
FMMP&9.1$\times10^{-4}$&1.4$\times10^{-1}$&2.9$\times10^{-2}$
&$1.7\times10^{-1}$&$0.14\times10^{-2}$\\
program based on our method&2.2$\times10^{-4}$&8.5$\times10^{-2}$
&9.0$\times10^{-3}$&$9.4\times10^{-2}$&$0.18\times10^{-2}$\\
\hline
\\
\end{tabular}
\begin{tablenotes}
\item[1] time for upward pass
\item[2] time for downward pass
\item[3] time for computing potential field and force field
\item[4] averaged values of the relative errors (average of $\sum\frac{PF_{FMM}-PF_{direct}}{PF_{direct}}$)
\end{tablenotes}
\end{threeparttable}
\end{table}
\begin{table}[H]
\captionsetup[table]{justification=raggedright, singlelinecheck=off}
\begin{threeparttable}
\caption{\label{tab:table2}{\bf number of particles=200000, and maximum level $l$=3}} 

\begin{tabular}{ccccccc}
\hline
program 
&\verb|t_up|
&\verb|t_down|
&\verb|t_pfs|&total (sec)&accuracy\\
\hline
\hline
FMMP&9.4$\times10^{-4}$&1.7$\times10^{-1}$&7.3$\times10^{-2}$
&$2.4\times10^{-1}$&$0.35\times10^{-2}$\\
program based on our method&2.3$\times10^{-4}$&9.1$\times10^{-2}$
&2.3$\times10^{-2}$&$1.1\times10^{-1}$&$0.40\times10^{-2}$\\

\hline
FMMP({\small maximum order}\\
{\small of multipoles is four})&5.3$\times10^{-4}$&8.1$\times10^{-2}$
&5.6$\times10^{-2}$&$1.4\times10^{-1}$&$1.1\times10^{-2}$\\
\hline
\\
\end{tabular}
\end{threeparttable}
\end{table}
\begin{table}[H]
\captionsetup[table]{justification=raggedright, singlelinecheck=off}
\begin{threeparttable}
\caption{\label{tab:table3}
{\bf number of particles=80000, and maximum level $l$=2}}
\

\begin{tabular}{ccccccc}
\hline
program 
&\verb|t_up|
&\verb|t_down|
&\verb|t_pfs|&total (sec)&accuracy\\
\hline
\hline
FMMP&1.9$\times10^{-4}$&9.7$\times10^{-3}$
&2.9$\times10^{-2}$&$3.9\times10^{-2}$&$0.093\times10^{-2}$\\
program based on our method&7.8$\times10^{-5}$&6.4$\times10^{-3}$&8.0$\times10^{-3}$
&$1.4\times10^{-2}$&$0.15\times10^{-2}$\\
\hline

\\
\end{tabular}
\end{threeparttable}
\end{table}

\section{Discussions}
(1) We first replaced spherical harmonics by charges fixed in position using
Lagrange interpolation. 
However, there may be other interpolation suitable for FMM.
Such a method had been
already invented by William Fong and Eric Darve \cite{120}. They used Chebyshev polynomials
to interpolate.
Their accuracies seem to be better than ours. However, since I have already coded my program 
before I knew 
their results and this paper aims to introduce a method that can remove "shift" in FMM,
I left as it was. 
It seems likely that the method introduced in this paper can be applied to Chebyshev polynomials.
\\
\\
(2) It takes time to compute the bilinear forms $B_{m_i,m_j}(v_k,v_l)$. However, since it does not depend
on $v_k$ and $v_l$, we can compute and store the bilinear forms in the memory in advance.
It requires only once, so I ignored the time required to compute the bilinear forms.
For the case of FMMP, some computation may be done in advance and 
shorten the computation time. 
\\
\\
(3) The bilinear forms $B_{m_i,m_j}(v_k,v_l)$ are essentially matrices. The computations to obtain
the matrices go through computation with large numbers and end up with relatively small number entries.
Thus, the computation of $B_{m_i,m_j}(v_k,v_l)$ should {\it not} be done in a single precision.
\\
\\
(4) As shown in the tables in the previous section, we could save computation time substantially (about half).
However, the accuracy of our method is slightly worse than that obtained by the FMMP  program.
It may become more accurate if we increase the dimension $\mathtt{d}$  of $\bf{V}$ or 
employ the method of William Fong and Eric Darve \cite{120}.
We find that if we set the maximum order of multipoles equal to four in FMM, the accuracy 
becomes considerably worse ($> 1\%$), but not so fast. 
\\
\\
(5) Since the array data $B_{m_i,m_j}(v_k,v_l)$ is very large, the time required to access memory
storing the array is a crucial factor to determine computation time. 
If we find a good way to manage the memory, it is probable that the 
computation time would become less.

\appendix
\section{Appendix}
In this appendix, I will give a construction of the bilinear forms $B_{m_i,m_j}(v_k,v_l)$ 
introduced in $\S2.2.1$ in more detail in the following steps. 
I also provide here descriptions for $\rm{\,I\,}$ and $\rm{I\hspace{-.01em}I}$ in $\S2.2.2$.
We assume that 
the dimension $\mathtt{d}$ of $\bf{V}$ is equal to 16
for the sake of description. We denote by $\phi(c,c\prime)$ 
a function determined by charges $c$ and $c\prime$ such as
Coulomb potential. We also write $a\thickapprox b$ if $a-b$ is within a predictable error.

\begin{enumerate}[{STEP}1]
\item We define 16 (=$\mathtt{d}$) positions in each square at every level (Fig.3).
\item Let a charge $c \in m^k$ with $k\ge 0$. We define a map $h_m$ from $c$ to
16 point-charges on the fixed 16 positions in $m$. 
For $k=0$ this map is the map introduced in $\S 2.2.1$.
The map $h_m$ has the following property (for detail, see A.2.2):
If charges $c_1 \in m_1^k$ and $c_2 \in m_2^k$ with $k\ge 1$ and $m_1$ and
$m_2$ are well-separated, then
\[
\phi(c_1,c_2)\thickapprox\sum^{16}_i \phi(c_1^i,c_2) 
\]
where $h_{m_1}(c_1)$=\{$c_1^i$, ($1\le i \le16$)\}.
By assigning the electric charges of the 16 point charges to the coordinate of 16 dimensional vector space, 
the image of the map $h_{m_1}$ can be regarded as a column vector in a 16-dimensional space.
We also denote this column vector by $h_{m_1}(c)$. We identify the vector space $\bf{V}$ with
the values of the 16 point-charges in $m^0$.

\item We define a bilinear form $b_{m_1,m_2}(v_1,v_2)\thickapprox\phi(c_1,c_2)$ 
for $v_1 \in m_1$ and $v_2 \in m_2$ by
\[
b_{m_1,m_2}(v_1,v_2)=\sum^{16}_i \sum^{16}_j\phi(c_1^i,c_2^j) 
\]
where $c_1\rightarrow v_1$ and $c_2\rightarrow v_2$,
$h_{m_1}(c_1)$=\{$c_1^i$, ($1\le i \le16$)\}, and $h_{m_2}(c_2)$=\{$c_2^j$, ($1\le j \le16$)\}.
\item By the STEP2 above, we can map a charge $c\in m$ to
 16 point charges $h_m(c)$ $\in m$.
Let $p_i\in m^0$ $(1\le i\le16)$ be the 16 point charges of $m^0$ arranged as in Fig.3.
We assume $Chg(p_i)=1$ $(1\le i\le16)$, where 
$Chg(p)$ denote the electric charge of $p$. 
We can transform each of the 16 point charges $p_i\in m^0$ in $m^0$ to the
16 point charges in $m^k_i$ by $h_{m^k_i}$.
Then, we define a $16\times16$ matrix $M_{m_1}$ by 
\[
M_{m_1}=(h_{m_i}(p_1),h_{m_i}(p_2),,,h_{m_i}(p_{16})).
\]
$M_{m_1}$ transforms the 16 charges of $m^0$ to the 16 charges in $m_1$.
\item We define bilinear forms $B_{m_i,m_j}(v_k,v_l)$ by
\[
B_{m_i,m_j}(v_k,v_l)=b_{m_i,m_j}(M_{m_i}v_k,M_{m_j}v_2).
\]
Writing $b_{m_1,m_2}(v_1,v_2)=v_1^TM_bv_2$ with a suitable matrix $M_b$, 
the matrix for $B_{m_i,m_j}(v_k,v_l)$ is
\[
M_{m_i}^TM_bM_{m_j},
\]
where $T$ denotes its transpose.
\end{enumerate}

In the following, I describe the steps above in more detail.

\subsection{STEP1}
We fix 16 points as illustrated in Fig.3. The number of points should be $n^2$ and $n^3$ for 
2-dimensional case and 3-dimensional case, respectively.

\begin{figure}[h]
\centering
\label{Fig.3}
\includegraphics[width=4cm]{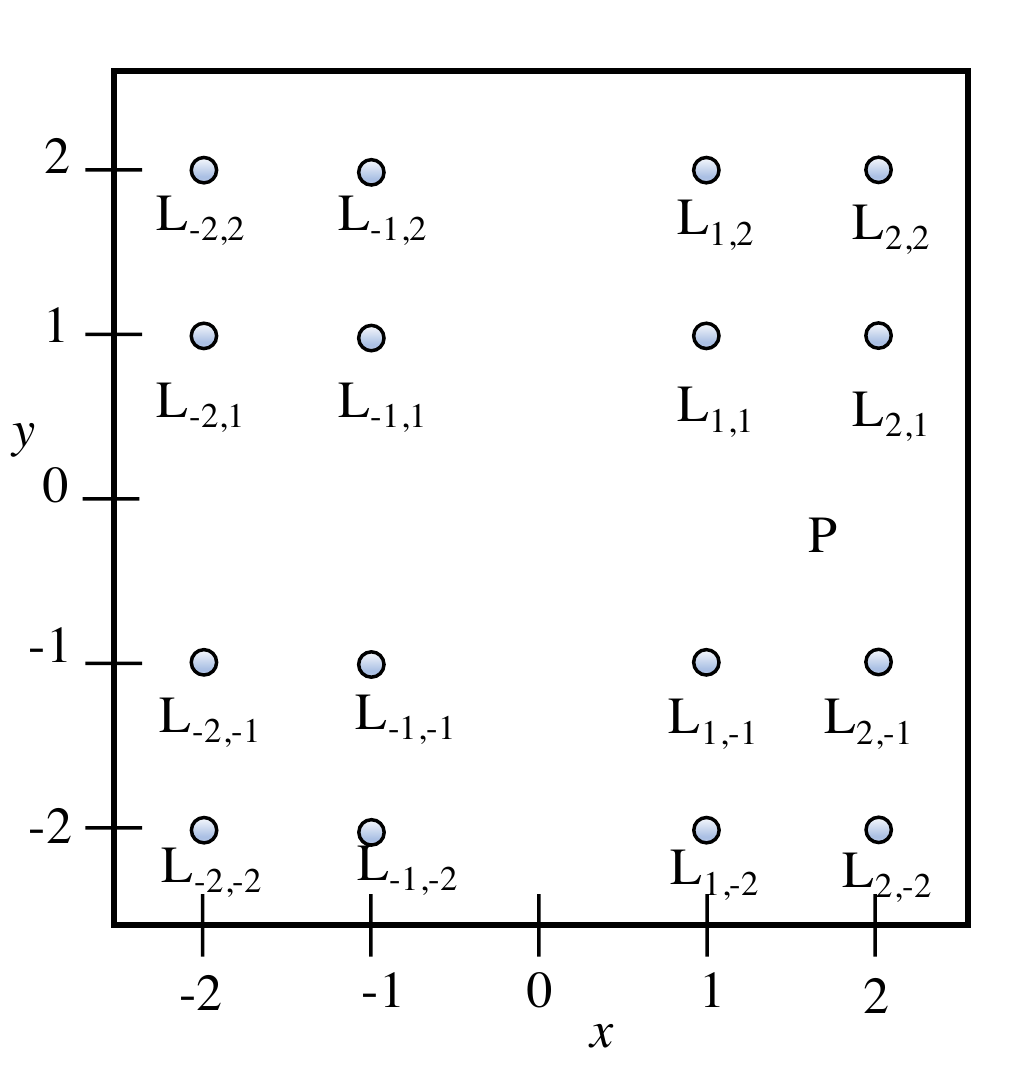}
\caption{16 points in a square. We can arbitrarily choose the number of points as long as it is a
squared integer. For the case of three-dimensional cases, it should be cubed integer.
We can choose other arrangements of points than the above.}
\end{figure}

\subsection{STEP2}
\subsubsection{Lagrange interpolation}
For any polynomial $g(x)$ with $\deg(g)<n$, we have
\begin{equation}
\sum^n_i f_i(x)g_i(a_i)-g(x)=0
\end{equation}
for arbitrarily chosen distinct $m$-numbers \{$a_i$\}, where 
$f_i(x)=\frac{F_i(x)}{F_i(a_i)}, \ F_k(x)=\prod^n_{i\ne k}(x-a_i)=\frac{\prod^n_{i=1}(x-a_i)}{(x-a_k)}$.

If $deg(g(x))\ge n$ or $g(x)$ is a power series, 
$\sum^n_j f_j(x)g(a_j)-g(x)=x^n\times({\rm power \ series})$.
Thus potential field is approximated with a predictable error.
\subsubsection{Approximation of a charge by charges fixed in position}

Let us assume that a charged particle is located at a point $P$. We represent the field of charge 
due to this particle at point  $A$ by four charges located at $P_2, P_1,  P_{-1}, P_{-2}$ by
applying the results above.
Let $\phi(P,A)$ be a function determined by charges $P$ and $A$.
For instance, $\phi(P,A)=\frac{Chg(P)\times Chg(A)}{|P-A|}$ where $Chg(X)$ denote the 
electric charge of the point charge $X$.

Let $A, \ B$ be points shown in Fig.4, $\phi(P,A)$ can be written as
\begin{equation}
\phi(P,A)\thickapprox\sum_{i,j\in\{2,1,-1,-2\}} f_i(x)f_j(y)\phi(L_{i,j},A)
\end{equation}
where $f_i$ is the function defined in the previous section for $n=4$ and $a_i=\{2,1,-1,-2\}$,
and $(x,y)$ is the coordinate of the point $P$ as depicted in the figure.
Putting $P=c_1$, $A=c_2$, and 
$\{c_1^k, \ 1\le k\le16 \}=\{f_i(x)f_j(y)Chg(P), \ 1\le i\le4, \ 1\le j\le4\}$
as $h_{m_1}(c_1)$=\{$c_1^i$, ($1\le i \le16$)\} described in STEP2, we get the relation
$\phi(c_1,c_2)\thickapprox\sum^{16}_i \phi(c_1^i,c_2)$.

\begin{figure}[h]
\centering
\label{Fig.4}
\includegraphics[width=4cm]{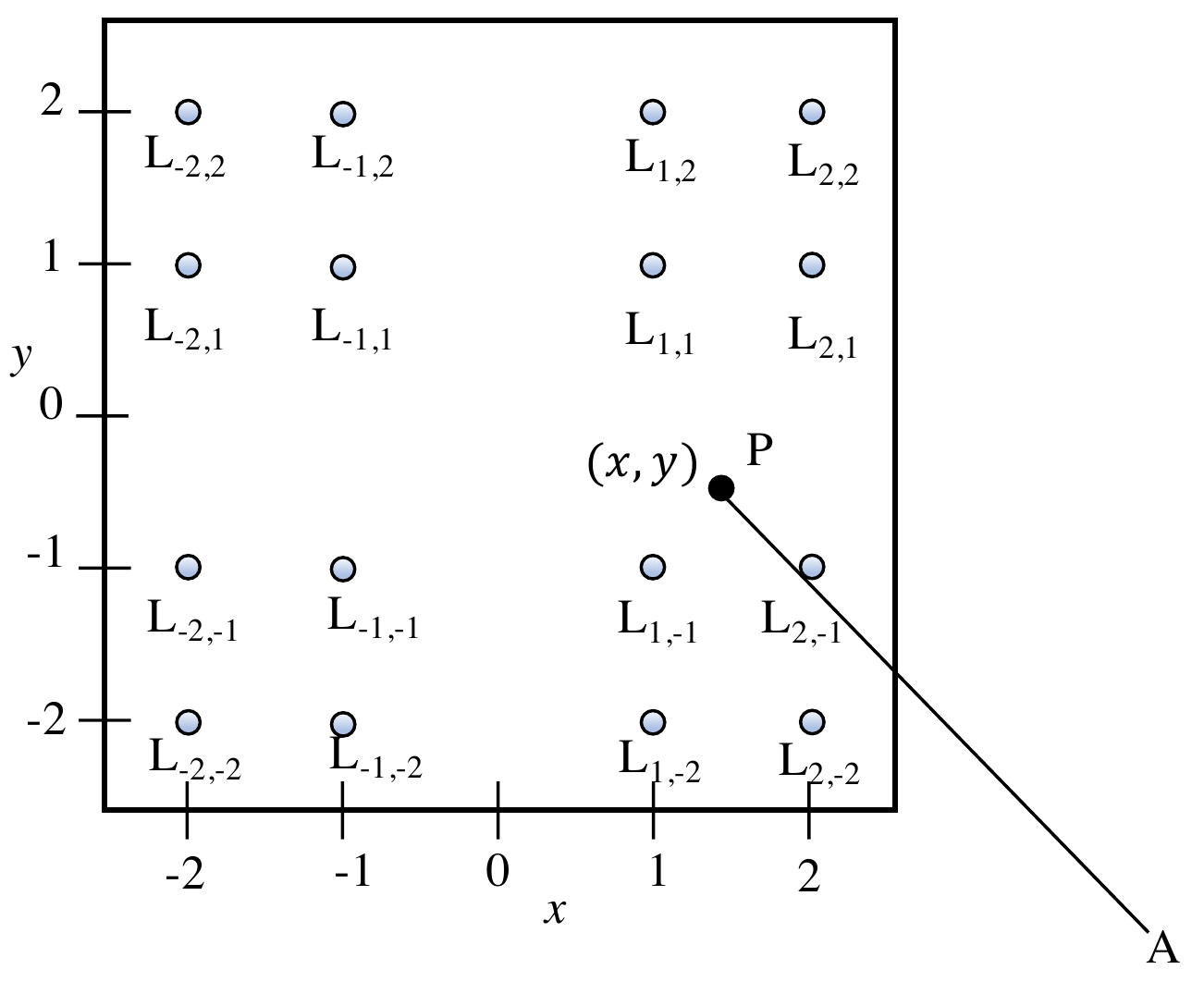}
\caption{Points A and B}
\end{figure}


\begin{thebibliography}{200}
\bibitem{101} Leslie F. Greengard, The Rapid Evaluation of Potential Fields in Particle Systems, The MIT Press, Cambridge, Massachusetts, 1988
\bibitem{110} The Best of the 20th Century: Editors Name Top 10 Algorithms, SIAM News, {\bf 33}, 1 (2000)
\bibitem{120} William Fong, Eric Darve, The black-box fast multipole method, 
J. Comp. Phys. {\bf 228}, 8712-8725 (2009)
\bibitem{121} Board, J. and Schulten, K. ,The fast multipole algorithm, Comput. Sci. Eng. {\bf 2}, 76-79 (2000)
\bibitem{122} C.A. White and M. Head-Gordon, Derivation and efficient implementation of the fast multipole method, J. Chem. Phys., {\bf 101}, 6593-6605 (1994)
\bibitem{123} H. Fujiwara, The fast multipole method for solving integral equations of three-dimensional topography and basin problems, Geophys. Int. J.,{\bf 140}, 198-210 (2000)
\bibitem{124} T. Hrycak and V. Rokhlin, An improved fast multipole algorithm for potential fields,SIAM J. Sci. Comput., {\bf 19}, 1804-1826 (1998)
\bibitem{140} Y. Kajima, S. Ogata, R. Kobayashi, M. Hiyama, and T. Tamura, Fluctuating Local Recrystallization of Quasi-Liquid Layer of Sub-Micrometer-Scale Ice: A Molecular Dynamics Study, J. Phys. Soc. Jpn. {\bf 83}, 83601 (2014)
\bibitem{130} Shuji Ogata, Timothy J. Campbell, Rajiv K. Kalia, Aiichiro Nakano, Priya Vashishta, and Satyavani Vemparala, Scalable and portable implementation of the fast multipole method on parallel computers, Comp. Phys. Comm., {\bf 153}, 445-461 (2003) 
\end{thebibliography}
\end{document}